Unterrichtsmaterialien zum
50-jährigen Jubiläum von Apollo 11

In Kooperation mit

**Handreichung für Lehrpersonen:**

# Wie brachte die Saturn V-Rakete die Astronauten von Apollo 11 zum Mond?

**Klassen 10 – 13**

Markus Nielbock

6. Mai 2019

## Zusammenfassung

Diese Lehreinheit schließt sich an das Material *Wie fliegen Astronauten mit einer Rakete zur ISS?* (http://www.haus-der-astronomie.de/raum-fuer-bildung) an, richtet sich aber wegen des benötigten mathematischen Rüstzeugs an höhere Klassenstufen. Die Schülerinnen und Schüler berechnen ausgehend von der Raketengleichung verschiedene Parameter der dreistufigen Mondrakete *Saturn V* unter dem Einfluss der Schwerkraft. Um die Berechnung analytisch durchführen zu können, werden einige vereinfachende Annahmen gemacht. Zur Motivation werden Videos und Texte zum Apollo-Programm zur Verfügung gestellt.

## Lernziele

Die Schülerinnen und Schüler

- machen sich mit dem Apollo-Programm vertraut,
- stellen das Flugprofil des Raketenstarts grafisch dar,
- berechnen die Geschwindigkeiten einer mehrstufigen Rakete mit realistischen Kennzahlen,
- ermitteln einige für den Raketenflug wichtige Parameter.

## Materialien

- Arbeitsblätter (erhältlich unter: http://www.haus-der-astronomie.de/raum-fuer-bildung)
- Stift
- Taschenrechner
- Mobiltelefon oder Computer mit Internetzugang

## Stichworte

Mond, Apollo 11, Saturn V, Rakete, Rückstoß, Impuls, Schub, Dichte

## Dauer

90 - 180 Minuten (je nach Auswahl der Aufgaben)

# Hintergrund

## Das Apollo-Programm

Das Apollo-Programm der NASA (National Aeronautics and Space Administration, USA) hatte das Ziel, innerhalb eines Jahrzehnts Menschen auf den Mond und wieder zurück zu bringen. Dieses Projekt war ein Resultat des Wettstreits der beiden Weltmächte, den USA und der UdSSR. Parallel zu den Erfolgen der Sowjetunion mit Sputnik[a] 1, dem ersten künstlichen Satelliten im Jahre 1957, sowie Juri Gagarin, der als erster Mensch 1961 ins All flog, etablierten auch die USA ihr eigenes Weltraumprogramm. Wegen des offensichtlichen Vorsprungs der UdSSR in der erdnahen Raumfahrt strebten die USA schließlich die Landung von Menschen auf dem Mond an, um noch Aussichten zu haben, die Sowjetunion in der Weltraumfahrt zu überholen (Dupas und Logsdon 1994).

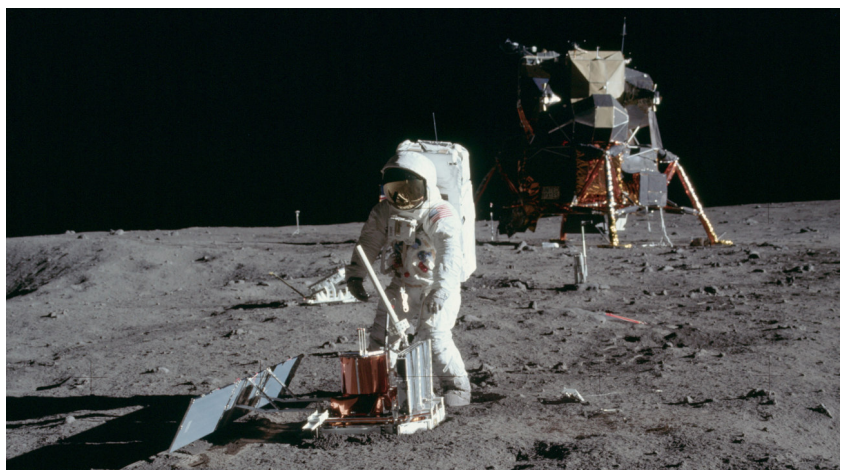

**Abbildung 1:** Die ersten Menschen landeten während der Apollo 11-Mission auf dem Mond. In diesem Bild nimmt Edwin Aldrin ein Seismometer auf dem Mond in Betrieb. Im Hintergrund steht das Landemodul (LM, Lunar Module) „Eagle" (Bild: NASA).

Von den 14 Flügen wurden drei (Apollo 4 bis 6) als Test ohne Besatzung durchgeführt. Neun von elf bemannten Apollo[b]-Missionen flogen zum Mond, wobei sechs auf dem Mond landeten. Die erste Landung erfolgte mit Apollo 11 am 20. Juli 1969 (Abb. 1); mit Apollo 17 erkundete die vorerst letzte Besatzung im Dezember 1972 unseren Trabanten.

## Saturn V-Rakete

Maßgeblich für den Erfolg des Apollo-Programms war die neu entwickelte Saturn V-Rakete (Abb. 2). Sie ist bis heute die leistungsfähigste Rakete, die je zum Einsatz kam. Die Rakete war darauf ausgelegt, Nutzlasten von bis zu 140 t in einen erdnahen Orbit (Rehmus 2006) und knapp 50 t bis zum Mond zu bringen. Die Saturn V wurde von Wernher von Braun (Lauer 2019) und Arthur Rudolph (Blumenthal 1984) am Marshall Space Flight Center in Huntsville, Alabama, USA entwickelt und von verschiedenen kommerziellen Partnern gebaut. Die dreistufige Rakete besaß eine Gesamthöhe von 111 m und einen Durchmesser von 10 m. Das Startgewicht betrug nahezu 3000 t (Orloff 2000, S. 284).

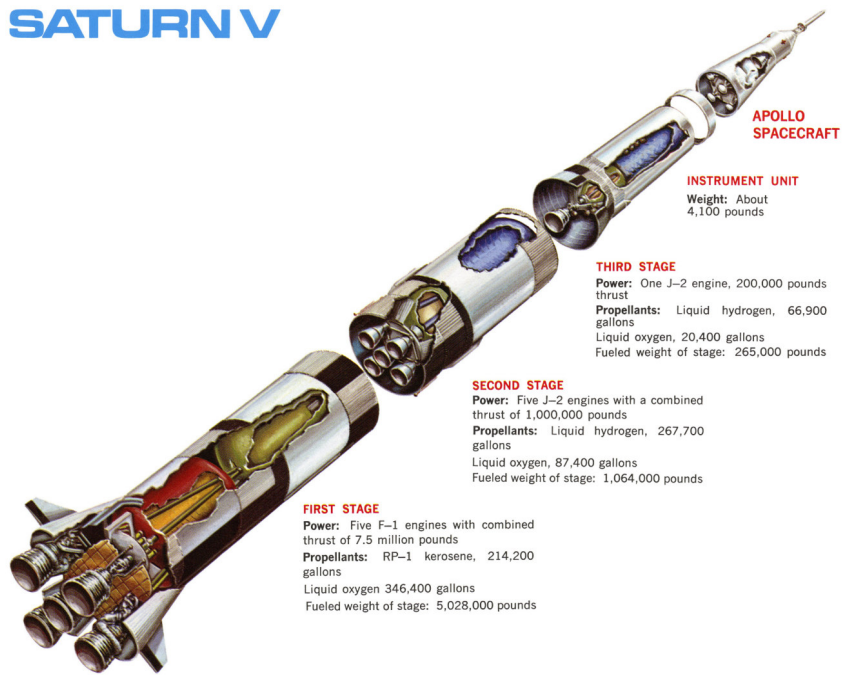

**Abbildung 2:** Grafische Darstellung der Hauptkomponenten der Saturn V-Rakete (Bild: Mike Jetzer/heroicrelics.org/NASA).

## Triebwerke und Stufen der Saturn V-Rakete

Das Konzept mehrstufiger Raketen ist schon seit mehreren Jahrhunderten bekannt (Barth 2005; Needham u. a. 1986, S. 485 ff.). Ihr Vorteil gegenüber einstufigen Raketen liegt darin, dass sie den Ballast aus leeren Tanks und somit überflüssigen Teilen der Rakete abwerfen. Dadurch können höhere Endgeschwindigkeiten erzielt werden als mit einer einstufigen Konfiguration.

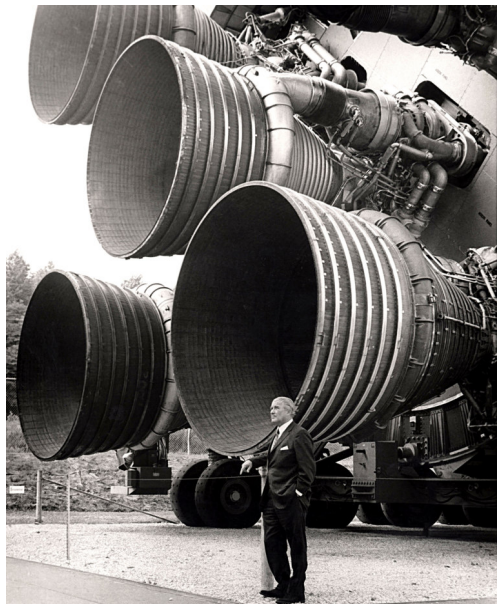

**Abbildung 3:** Wernher von Braun vor der ersten Stufe der Saturn-V-Rakete mit fünf F-1-Triebwerken (Bild: NASA).

Für die Saturn V-Rakete wurden zwei neue, sehr leistungsstarke Triebwerke von der Firma Rocketdyne entwickelt. Das F-1-Triebwerk, von dem die erste Stufe (S-IC) gleich fünf besaß, ist bis heute eins der stärksten Raketentriebwerke, die je gebaut wurden (Woods 2011, S. 21). Es verbrannte je Sekunde ca. 2,6 t eines Gemischs aus flüssigem Sauerstoff und Kerosin.

Die beiden anderen Raketenstufen waren mit J-2-Triebwerken bestückt, die ein Gemisch aus flüssigem Sauerstoff und flüssigem Wasserstoff verbrannten. Die zweite Stufe (S-II) wurde mit fünf J-2-Triebwerken betrieben, während die dritte Stufe (S-IVB) lediglich von einem beschleunigt wurde.

### Die Physik eines Raketenantriebs

Wir kennen bereits das Prinzip des freien Falls. Ein Objekt mit einer Masse $m$ wird durch die Gravitationskraft der Erde angezogen. Diese Kraft führt dazu, dass dieses Objekt – einmal losgelassen – auf die Erde fällt. Dabei nimmt die Geschwindigkeit stetig zu. Die zeitliche Änderung der Geschwindigkeit nennt man Beschleunigung. Dieser Zusammenhang ist auch als die *Grundgleichung der Mechanik* oder *2. Newtonsches Axiom* bekannt. Daher kann man schreiben:

$$\vec{F}_g = m \cdot \vec{a} \tag{1}$$

$$\text{mit: } \vec{a} = \frac{\mathrm{d}\vec{v}}{\mathrm{d}t} \tag{2}$$

Diese Gleichung setzt also die auf $m$ wirkende Kraft $\vec{F}_g$ in Beziehung zur Beschleunigung $\vec{a}$, die es erfährt. In Bodennähe kann die Kraft vereinfacht durch $m \cdot \vec{g}$ geschrieben werden, wobei $\vec{g}$ die Erdbeschleunigung ist. Wir nehmen hier den Wert am Äquator der Erde an ($|\vec{g}| = 9,8\,\mathrm{m/s^2}$).

$$\vec{F}_g = m \cdot \vec{g} = m \cdot \vec{a} \tag{3}$$

Mit Gl. 2 erhält man dann:

$$m \cdot \vec{g} = m \cdot \frac{\mathrm{d}\vec{v}}{\mathrm{d}t} \tag{4}$$

Die Geschwindigkeit der Masse $m$ nimmt daher im freien Fall je Zeiteinheit $\Delta t$ um $\Delta v$ zu, also in einer Sekunde um 9,8 m/s. Wir sehen hier auch, dass ohne weitere äußere Krafteinwirkung (z. B. Luftreibung) die Masse $m$ sich wegkürzt. Daraus folgt, dass der Geschwindigkeitszuwachs eines Objekts nicht von seiner Masse, sondern lediglich von der wirkenden Erdbeschleunigung abhängt.

Mit einer Rakete möchten wir das genaue Gegenteil erreichen, nämlich eine Nutzlast entgegen der wirkenden Gravitation nach oben befördern. Dazu muss Kraft wirken, die man als Schubkraft bezeichnet. Hierfür könnte man schlicht $\vec{F}_S$ schreiben. Es hat sich jedoch die Schreibweise $\vec{S}$ mit $|\vec{S}| = S$ etabliert. Diese Schubkraft, oder kurz Schub, wird durch den Auswurf von verbranntem Treibstoff unter hoher Geschwindigkeit $\vec{w}$ erzeugt. Man benutzt hier das Formelzeichen $w$, um die Geschwindigkeit der Triebwerkgase von der Geschwindigkeit der Rakete zu unterscheiden. Die Masse des Treibstoffs wird der Gesamtmasse der Rakete entzogen und wird deshalb mit $\Delta m$ bezeichnet.

Der Treibstoffdurchsatz $\mu = \Delta m / \Delta t$ zeigt also an, mit welcher Rate im Mittel Treibstoff verbraucht wird und sich die Masse der Rakete ändert.

Physikalisch entspricht eine Kraft einer zeitlichen Änderung des Impulses $\vec{p}$. Diese Größe beschreibt den Bewegungszustand eines Objekts. Allgemein lässt sich für den Schub einer Rakete also schreiben:

$$\vec{S} = \frac{\mathrm{d}\vec{p}}{\mathrm{d}t} \tag{5}$$

In diesem Fall ist $\vec{p}$ der Impuls des ausgestoßenen Verbrennungsprodukte des Triebwerks. Mit

$$\vec{p} = m \cdot \vec{w} \tag{6}$$

folgt somit:

$$\vec{S} = \frac{\mathrm{d}}{\mathrm{d}t}(m \cdot \vec{w}) \tag{7}$$

Gewöhnlich fungiert die Masse als eine Konstante der Trägheit, die die Geschwindigkeitsveränderung beeinflusst, so dass man sie ausklammern könnte. Jedoch ist das bei Triebwerken gerade nicht der Fall, da mit $\mu = \Delta m / \Delta t$ Masse ausgeworfen wird. Deswegen folgt mit der Produktregel:

$$\vec{S} = \frac{\mathrm{d}m}{\mathrm{d}t} \cdot \vec{w} + m \cdot \frac{\mathrm{d}\vec{w}}{\mathrm{d}t} \tag{8}$$

Nimmt man vereinfachend an, dass $\vec{w}$ und $\mu$ konstant sind, folgt:

$$\vec{S} = \frac{\mathrm{d}m}{\mathrm{d}t} \cdot \vec{w} = \frac{\Delta m}{\Delta t} \cdot \vec{w} = \mu \cdot \vec{w} \tag{9}$$

Die Einheit des Schubs entspricht der einer Kraft, somit: $\mathrm{kg \cdot m/s^2 = N}$. Um mit der Rakete abheben zu können, muss also vom Betrag her stets $S > F_g$ gelten, wobei die Masse der Rakete $m_R$ ständig pro $\Delta t$ um $\Delta m$ abnimmt. Die Fähigkeit einer Rakete, den Erdboden zu verlassen, hängt also von der Startmasse der Rakete, dem Treibstoffdurchsatz $\mu$ und der Ausströmgeschwindigkeit $w = |\vec{w}|$ ab. Die letzteren beiden Größen sind charakteristisch für die verschiedenen Triebwerke, die in der Raumfahrt benutzt werden.

**Der spezifische Impuls**

In der Raketentechnik hat sich der Begriff des *spezifischen Impulses* $I_{\mathrm{sp}}$ eingebürgert. Hierunter versteht man den Impuls pro Massenelement des von einem Triebwerk ausgestoßenen Verbrennungsprodukts. Definiert ist er als das Produkt des über die Brenndauer $\Delta t$ gemittelten Schubs $\bar{S}$ und der Brenndauer geteilt durch die Masse des verbrannten Treibstoffs $\Delta m$. Das Produkt aus dem Schub und der Brenndauer ist lediglich der zeitlich gemittelte Impuls der ausströmenden Gases.

$$I_{\mathrm{sp}} = \frac{\bar{S} \cdot \Delta t}{\Delta m} \tag{10}$$

$$= \frac{\bar{p}}{\Delta m} \tag{11}$$

Die Einheit von $I_{\mathrm{sp}}$ ist somit m/s, also eine Geschwindigkeit. $I_{\mathrm{sp}}$ ist daher bis auf technische Verluste (z. B. Reibung, Brennkammereffizienz) die Ausströmgeschwindigkeit $w$, die sich jedoch in der Praxis zeitlich ändert. Das erkennt man auch durch den Vergleich mit der Definition des Schubs in Gl. 9. Sowohl der Schub als auch der spezifische Impuls sind von den äußeren Druckbedingungen abhängig, da das Triebwerk gegen diesen Druck arbeitet. Deswegen steigen typischerweise $\bar{S}$, $I_{sp}$ und $w$ mit zunehmender Höhe und abnehmendem atmosphärischem Druck. Im folgenden nehmen wir jedoch an, dass $S$, $I_{sp}$ und $w$ zeitlich konstant sind.

Oft bezieht man den spezifischen Impuls nicht auf die Masse des Gases, sondern auf die Gewichtskraft unter der Einwirkung der Erdbeschleunigung $g$ (Messerschmid und Fasoulas 2011). Die resultierende Einheit entspricht dann derjenigen der Zeit. Wir werden jedoch stets mit der Definition nach Gl. 10 arbeiten.

$$I_{sp}^{\star} = \frac{\bar{S} \cdot \Delta t}{\Delta m \cdot g} \tag{12}$$

**Raketengleichung**

Für die weitere Betrachtung wird die vektorielle Schreibweise zugunsten einer betragsmäßigen Darstellung der Größen vereinfacht, da in der idealisierten Beschreibung die Richtungen der Geschwindigkeiten und Kräfte stets parallel bzw. antiparallel sind.

Die sogenannte Raketengleichung oder auch Ziolkowski-Gleichung – benannt nach dem russischen Weltraumpionier Konstantin Ziolkowski, der diese Grundgleichung 1903 aufstellte (Lossau 2010) – beschreibt die Bewegung einer einstufigen, kräftefreien Rakete. Sie lässt sich über zwei äquivalente Prinzipien herleiten – die Grundgleichung der Mechanik bzw. das 2. Newtonschen Axiom und die Impulserhaltung.

Aus der Grundgleichung der Mechanik lässt sich eine Bewegungsgleichung für das Verhalten einer Rakete mit der Masse $m_R$ und dem Schub $S$ nach Gl. 9 aufstellen. Sie erfährt die Beschleunigung $a_R$.

$$F = m \cdot a \Leftrightarrow S = m_R \cdot a_R \tag{13}$$

In diesem Fall ist $m_R$ von der Zeit $t$ abhängig, denn mit der Zündung der Rakete verbrennt Treibstoff mit dem Durchsatz $\mu$. Somit kann man schreiben:

$$m_R(t) = m_0 - \mu \cdot t \tag{14}$$

Dabei ist $m_0$ die Masse der Rakete bei Brennbeginn. Die Beschleunigung nimmt im Laufe der Zeit also zu. Mit der Annahme, dass $S = \mu \cdot t$ konstant ist (was in der Realität nicht stimmt), kann man schreiben:

$$\mu \cdot w = (m_0 - \mu \cdot t) \cdot a_R(t) \tag{15}$$

$$\Leftrightarrow a_R(t) = \frac{\mu \cdot w}{m_0 - \mu \cdot t} \tag{16}$$

Für kleine Veränderungen von $\Delta t$ können wir Gl. (15) näherungsweise schreiben:

$$\frac{\Delta m}{\Delta t} \cdot w = (m_0 - \Delta m) \cdot \frac{\Delta v_R}{\Delta t} \tag{17}$$

$$\Leftrightarrow \Delta m \cdot w = (m_0 - \Delta m) \cdot \Delta v_R \tag{18}$$

$$\Leftrightarrow \Delta v_R = \frac{\Delta m}{m_0 - \Delta m} \cdot w \tag{19}$$

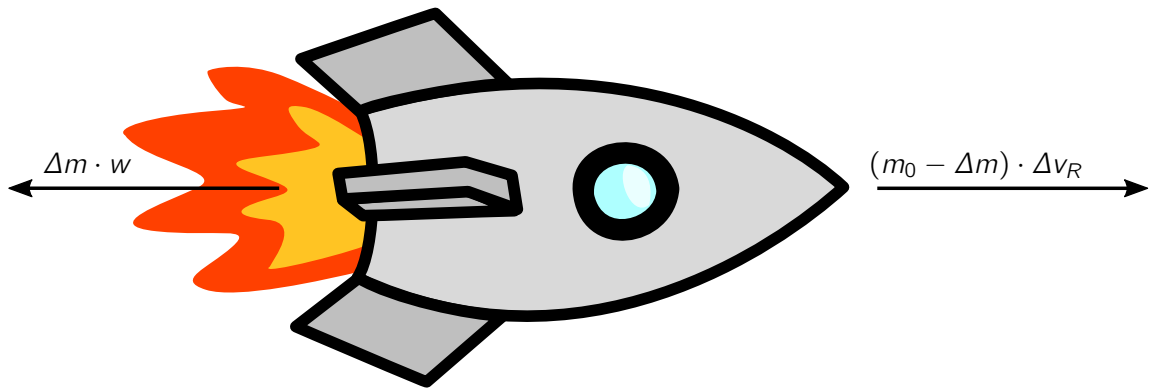

In Gl. 18 erkennt man die Impulserhaltung mit den Teilimpulsen des ausströmenden Gases und des Raketenkörpers. Für genügend kleine $\Delta m = \mu \cdot \Delta t$ kann man mit Hilfe von Gl. 19 die Geschwindigkeitsänderung der Rakete für bekannte Werte von $\mu$ und $w$ berechnen. Das ist besonders hilfreich, wenn man die Endgeschwindigkeit einer Raketenstufe mit der Methode der kleinen Schritte annähern möchte (siehe: *Wie fliegen Astronauten mit einer Rakete zur ISS?*, http://www.haus-der-astronomie.de/raum-fuer-bildung).

Tatsächlich muss man für das korrekte Ergebnis ein Integral lösen, wobei man von Gl. 16 ausgehend $v(t) = \int a_R(t)\,dt$ auswertet. Wir nehmen an, dass $\mu$ und $w$ konstant sind.

$$v_R(t) = \int_{t_0}^{t_1} a_R(t)\,dt = \int_{t_0}^{t_1} \frac{\mu \cdot w}{m_0 - \mu \cdot t}\,dt = -w \cdot \int_{t_0}^{t_1} \frac{-\mu}{-\mu \cdot t + m_0}\,dt \qquad (20)$$

Mit $\frac{d}{dx}\ln(a \cdot x + b) = \frac{a}{a \cdot x + b}$ folgt:

$$\begin{aligned}
[v_R(t)]_{t_0}^{t_1} &= -w \cdot [\ln(m_0 - \mu \cdot t)]_{t_0}^{t_1} \\
\Leftrightarrow v_R(t_1) - v_R(t_0) = \Delta v_R &= -w\,[\ln(m_0 - \mu \cdot t_1) - \ln(m_0 - \mu \cdot t_0)] \\
\Leftrightarrow \Delta v_R &\stackrel{\substack{t_0=0\\\mu \cdot t_1 = \Delta m}}{=} -w \cdot [\ln(m_0 - \Delta m) - \ln(m_0)] \\
\Leftrightarrow \Delta v_R &\stackrel{m_B = m_0 - \Delta m}{=} -w \cdot \ln\left(\frac{m_B}{m_0}\right) \\
\Leftrightarrow \Delta v_R &= w \cdot \ln\left(\frac{m_0}{m_B}\right) \qquad (21)
\end{aligned}$$

Hier bedeuten $m_0$ die Gesamtmasse der Rakete zu Brennbeginn und $m_B$ die Masse der Rakete bei Brennende. Daraus ersieht man, dass eine Änderung der Geschwindigkeit vom natürlichen Logarithmus des Massenverhältnisses zu Beginn und Ende der Triebwerkzündung abhängt. Damit muss für eine bestimmte Geschwindigkeitsänderung eine überproportional zusätzlich große Treibstoffmenge umgesetzt werden, was zu einer deutlich größeren Rakete mit einer deutlich höheren Masse führt. Beispiel: Wir nehmen an, dass die Gesamtmasse einer Rakete zu vier Teilen vom Treibstoff und zu einem Teil von der Rakete selbst stammt. Will man für solch eine Rakete die Geschwindigkeit verdoppeln, benötigt man sechs mal so viel Treibstoff.

**Senkrecht aufsteigende Rakete**

Bislang wurde das Verhalten einer Rakete ohne äußere Kräfte betrachtet. Soll eine Rakete eine Nutzlast in einen Erdorbit oder zu einem anderen Himmelskörper bringen, arbeitet sie gegen die Gravitationswirkung der Erde an. Im folgenden wird vereinfacht angenommen, dass die Rakete senkrecht in die Höhe startet, also Gravitation und Schub in entgegengesetzte Richtungen wirken. In der Realität ist das nicht so, denn die Raketen verändern mit zunehmender Höhe ihre Neigung, um schließlich in einen Orbit parallel zur Erdoberfläche einzumünden. Dabei wird die Rotation der Erde ausgenutzt, deren Geschwindigkeit an der Erdoberfläche sich zu der finalen Bahngeschwindigkeit der Rakete hinzu addiert.

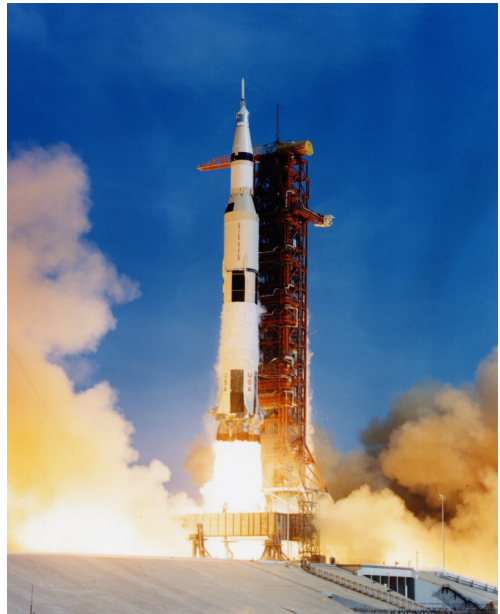

**Abbildung 4:** Die Saturn V-Rakete der Apollo 11-Mission wenige Sekunden nach dem Abheben von der Startrampe (Bild: NASA).

Insgesamt würde eine realistische Betrachtung eines Raketenstarts den Schwierigkeitsgrad für eine Behandlung im Schulunterricht übersteigen. Tatsächlich werden die Trajektorien von Raketen nicht analytisch, sondern rein numerisch mit mathematischen Modellen berechnet. Weiterhin benutzen wir für die Gewichtskraft $F_g = m \cdot g$. Zwar nimmt die Kraft mit zunehmendem Abstand von der Erde ab. Der resultierende Fehler ist aber für dieses Material vernachlässigbar.

Mit Gl. (3) folgt für $m_R(t)$:

$$F_g = m_R(t) \cdot g = (m_0 - \mu \cdot t) \cdot g \qquad (22)$$

Mit der bereits erwähnten der Annahme, dass die Gewichtskraft der Schubkraft genau entgegen wirkt, folgt, dass der effektive Schub $S^\star$ der um die Gewichtskraft $F_g$ verminderte Schub der Rakete $S$ ist. Somit:

$$S^\star = S - F_g = \mu \cdot t - (m_0 - \mu \cdot t) \cdot g \qquad (23)$$

Gleichzeitig gilt, dass der effektive Schub $S^\star$ zu einer Beschleunigung $a_R(t)$ der Rakete führt, die gegenüber der Betrachtung ohne den Einfluss der Schwerkraft geringer ist (siehe Gl. 13).

$$S^\star = m_R(t) \cdot a_R(t) = (m_0 - \mu \cdot t) \cdot a_R(t) \tag{24}$$

Daraus folgt:

$$a_R(t) = \frac{S^\star}{m_0 - \mu \cdot t} = \frac{\mu \cdot t - (m_0 - \mu \cdot t) \cdot g}{m_0 - \mu \cdot t}$$

$$\Leftrightarrow a_R(t) = \frac{\mu \cdot t}{m_0 - \mu \cdot t} - g \tag{25}$$

Dies entspricht also exakt der Beschleunigung einer kräftefreien Rakete (Gl. 16) vermindert um die Erdbeschleunigung $g$. Mit einer Rechnung äquivalent zu Gl. (20) folgt die Geschwindigkeitsänderung für die Brenndauer $\Delta t$.

$$\Delta v_R = w \cdot \ln\left(\frac{m_0}{m_B}\right) - g \cdot \Delta t \tag{26}$$

Wie in Gl. (21), sind auch hier $m_0$ die Masse der Rakete bei Brennbeginn und $m_B = m_0 - \Delta m$ die Masse bei Brennende.

**Mehrstufige Raketen**

Mehrstufige Raketen bestehen im Grunde aus mehreren einzelnen Raketen. Die erste Stufe bringt den Rest der Rakete auf eine gewisse Geschwindigkeit und Höhe. Nach der Trennung der ersten Stufe setzt die restliche Rakete den Flug aus eigenem Antrieb fort. Da der Ballast der leeren ersten Stufe fehlt, führt die geringere Masse zu einer höheren Beschleunigung und höheren Endgeschwindigkeit im Vergleich zu einer einstufigen Rakete.

Um die Geschwindigkeit einer mehrstufigen Rakete zu berechnen, kann man Gl. (26) für jede Stufe getrennt anwenden und die jeweiligen Geschwindigkeiten addieren. Somit folgt theoretisch für eine Rakete mit $n$ Stufen:

$$v_R = \sum_{i=1}^{n} \Delta v_{R,i} = \sum_{i=1}^{n} w_i \cdot \ln\left(\frac{m_{0,i}}{m_{B,i}}\right) - g \cdot \Delta t_i \tag{27}$$

Bei der Berechnung muss man beachten, dass die Masse der Rakete genommen wird, und nicht die Masse der jeweiligen Stufe. Die Masse der Rakete beim Brennbeginn der zweiten Stufe $m_{0,2}$ ist die Masse der Rakete bei Brennende der ersten Stufe $m_{B,1}$ abzüglich der Leermasse der ersten Stufe.

## Aktivität: Raketenflug

### Vorbereitung für Lehrpersonen

Lesen Sie das Kapitel mit den Hintergrundinformationen sorgfältig. Zusätzliche Literatur finden Sie am Ende dieses Dokuments.

Die Schülerinnen und Schüler benötigen für diese Aktivität bereits Vorkenntnisse zu dem Grundlagen der Mechanik, d. h. Grundgleichung der Mechanik, Begriff des Impulses, Begriff der Kraft, sowie aus der Mathematik Kenntnisse zur Differentialrechnung.

Machen Sie sich mit den Aufgaben der Schülerinnen und Schüler vertraut. Fertigen Sie ausreichend Kopien der Arbeitsblätter an. Für die Aktivität benötigen Sie Taschenrechner.

Dieses Material beinhaltet wegen des großen Umfangs des Themengebiets „Raketenflug" lediglich eine Einführung in die Grundbegriffe der Raketenantriebstechnik sowie eine teilrealistische Berechnung der Endgeschwindigkeit einer mehrstufigen Rakete nach dem Umsetzen des Treibstoffs in kinetische Energie. Hierzu werden vereinfachte Annahmen gemacht. Wir betrachten eine Rakete, die senkrecht aufsteigt und gegen die Erdgravitation arbeitet.

### Thematische Einführung (Vorschlag)

Zur Einführung in das Thema bietet es sich an, die Schülerinnen und Schüler zu ihren bisherigen Kenntnissen und Erfahrungen zur Raumfahrt zu befragen. Fragen Sie sie, ob sie einen (deutschen) Astronauten kennen und was sie über ihn wissen. Als thematische Einführung können zudem Fotos und kurze Videos von der Internationalen Raumstation gezeigt werden. So gibt es faszinierende Livestreams von der ISS.

https://www.nasa.gov/multimedia/nasatv/iss_ustream.html

http://www.ustream.tv/channel/iss-hdev-payload

Fragen Sie die Schülerinnen und Schüler, was Sie über die Mondlandungen und das Apollo-Programm wissen. Sie sollten versuchen, die Diskussion auf die verwendete Rakete, die Saturn V, zu lenken.

Informationen und Videos zur Einstimmung auf das Apolloprogramm helfen bei der Annäherung an das Thema.

DLR_next – Apollo 11 – Mond Special
www.dlr.de/next-apollo

DLR: Best of Apollo - 50 Jahre Mondlandung
https://youtu.be/8iNq_S8O7K4 (3:41 min)

Falls Sie ausreichend Zeit haben, lassen Sie die Lernenden im Internet über Apollo, insbesondere Apollo 11, und die Saturn V-Rakete recherchieren. Die Thematik lässt sich sehr gut durch Referate zu verschiedenen Themen der bemannten Raumfahrt innerhalb eines längeren Projekts behandeln.

Einen guten Überblick über die verschiedenen Phasen einer Apollo-Mission ermöglichen die folgenden drei Animationsvideos. Sie wurden in englischer Sprache verfasst, beinhalten jedoch deutsche Untertitel.

Wie das Apollo Raumschiff funktioniert: Teile 1 - 3
https://youtu.be/8dpkmUjJ8xU (3:57 min, Englisch)
https://youtu.be/tl1KPjxKVqk (5:17 min, Englisch)
https://youtu.be/qt_xoCXLXnI (4:01 min, Englisch)

Die originale Fernsehübertragung des Starts von Apollo 11 vom 16. Juli 1969 ist im folgenden Video zu sehen.

1969 Apollo 11 Saturn V launch, 1969 TV broadcast
https://youtu.be/xdxzMPi19sU (38:39 min, Englisch)

Die originalen Fernsehbilder der Apollo 11-Mission vom Mond werden von der NASA unter folgendem Link angeboten.

Apollo 11 HD Videos
https://www.nasa.gov/multimedia/hd/apollo11_hdpage.html

Alle Fotos, die jemals während des Apollo-Programms gemacht wurden, sind im Project Apollo Archive gesammelt und sind für jeden frei verfügbar.

Project Apollo Archive
https://www.flickr.com/photos/projectapolloarchive/albums

Mit etwas technischem Geschick können die Lernenden selbst eine Wasserrakete bauen. Dies kann beispielsweise mit dem Fach Technik verknüpft werden oder im Rahmen einer AG als Gruppenprojekt durchgeführt werden. Es existiert eine Vielzahl von Bauanleitungen mit unterschiedlichen Schwierigkeitsgraden.

https://www.dlr.de/next/desktopdefault.aspx/tabid-6297/

http://www.raketfuedrockets.com

Abhängig von den mathematischen Vorkenntnissen bietet es sich an, gemeinsam mit den Lernenden die Raketengleichung (Gln.21 und 26) herzuleiten.

## Aufgaben

### „Kugelstoßen" in Schwerelosigkeit

Ein Astronaut hat sich durch Unachtsamkeit von seinem Raumschiff entfernt. Nun versucht er, wieder zu seinem Schiff zurück zu kommen. Er hat einen Akkuschrauber dabei, der eine Masse von 10 kg aufweist. Der Astronaut wirft ihn in die vom Schiff abgewandte Richtung. Berechnen Sie mit den folgenden Angaben und der Raketengleichung (vgl. Gl. 21)

$$\Delta v = w \cdot \ln\left(\frac{m_0}{m_B}\right)$$

den Schub $S = \mu \cdot w$ (Gl. 9) des resultierenden Antriebs und die Geschwindigkeit $\Delta v$, mit der der Astronaut sich seinem Raumschiff nähert.

| | |
|---|---|
| Masse Astronaut | 190 kg |
| Masse Akkuschrauber | 10 kg |
| Abwurfgeschwindigkeit | 10 m/s |
| Abwurfdauer | 0,25 s |

### Flugprofil der Saturn V-Rakete

Um Raumschiffe und ihre Besatzungen zum Mond zu bringen, wurden zwischen 1968 und 1972 Saturn V-Raketen genutzt. Allerdings fliegt solch eine Rakete nicht senkrecht zum Erdboden in einem direkten Weg zum Mond, sondern steuert zunächst einen erdnahen Orbit an – einen Parkorbit. Erst danach setzt das Raumschiff seine Reise zum Mond fort. Daher schwenkte auch die Saturn V zunächst von einer senkrechten Startposition während ihres Aufstiegs langsam in eine Bahn parallel zur Erdoberfläche ein.

Zeichnen Sie das Flugprofil aus den nachfolgenden Daten (Orloff 2000, S. 103, 284) in ein geeignetes Koordinatensystem. Die horizontale Achse entspricht der Strecke der Rakete über Grund, während die senkrechte Achse die Höhe der Rakete angibt. Die Größen sollen in Kilometern angegeben werden.

**Tabelle 1:** Zurückgelegte Strecke und Höhe der Saturn V-Rakete der Apollo 11-Mission während der Startphase (Orloff 2000, S. 103).

| Ereignis | Zeit seit Abheben (mm:ss,ss) | Strecke über Grund (km) | Höhe (km) |
|---|---|---|---|
| Abheben | 00:00,63 | 0,00 | 0,05 |
| Mach 1 erreicht | 01:06,30 | 1,71 | 6,93 |
| Höchster dynamischer Druck | 01:23,00 | 4,93 | 11,98 |
| Stufe 1: Abschalten des zentralen Triebwerks | 02:15,20 | 41,00 | 38,87 |
| Stufe 1: Abschalten der restlichen Triebwerke | 02:41,63 | 82,65 | 58,40 |
| Stufe 1: Abtrennung und Zündung Stufe 2 | 02:42,30 | 83,95 | 58,93 |
| Stufe 2: Abschalten des zentralen Triebwerks | 07:40,62 | 984,15 | 159,12 |
| Stufe 2: Abschalten der restlichen Triebwerke | 09:08,22 | 1429,39 | 165,44 |
| Stufe 2: Abtrennung und Zündung Stufe 3 | 09:09,00 | 1433,75 | 165,49 |
| Stufe 3: Abschalten des Triebwerks | 11:39,33 | 2325,86 | 168,80 |
| Einschwenken in erdnahen Orbit | 11:49,33 | 2389,22 | 168,76 |

Schätzen Sie die Höhe ab, ab der sich die Rakete merklich von einer vertikalen Flugbahn entfernt.

### Eigenschaften der ersten Stufe der Saturn V-Rakete

Die erste Stufe der Saturn V-Rakete, genannt S-IC, wurde mit fünf neuartigen und sehr leistungsstarken Triebwerken vom Typ F-1 versehen (siehe Abb. 3). Die Auswertung der Flugdaten hat ergeben, dass Kerosin (RP-1, Magee u. a. 2007) mit flüssigem Sauerstoff (LOX) in einem Mischungsverhältnis von 1:2,343 verbrannt wurde. Dabei wurden laut Telemetrie 1 481 439 kg LOX und 632 635 kg RP-1 während der gesamten Brenndauer von 168,03 s umgesetzt (Orloff 2000, S. 289).

Berechnen Sie den Treibstoffdurchsatz $\mu$ während der Brenndauer $\Delta t$.

Die Startmasse der Saturn V betrug 2 938 315 kg (ebd., S. 284). Die Ausströmgeschwindigkeit hatte einen Wert von $w = 3180$ m/s. Berechnen Sie die Endgeschwindigkeit bei Brennschluss der ersten Stufe unter Berücksichtigung der Gravitationswirkung der Erde (siehe Gl. 26).

Ermitteln Sie den Schub $S$ der ersten Stufe der Saturn V.

### Treibstoff der zweiten Stufe der Saturn V-Rakete

Die zweite Stufe, bezeichnet mit S-II, wurde mit flüssigem Sauerstoff und Wasserstoff betrieben, die während des Flugs in getrennten Tanks sicher aufbewahrt und einer kontrollierten Reaktion zugeführt wurden. Die Raketenstufe besaß Triebwerke vom Typ J-2. Nach einer Brenndauer von 384,22 s erreichte die Rakete eine Geschwindigkeit von 6513 m/s.

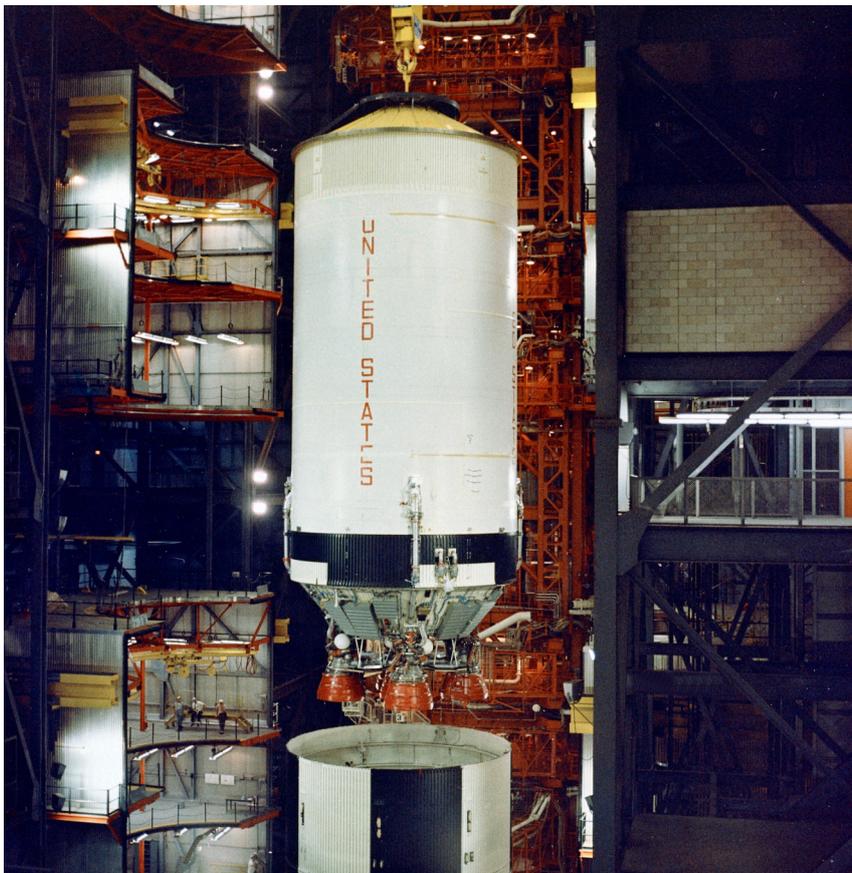

**Abbildung 5:** Foto der zweiten Stufe der Saturn V-Rakete mit ihren fünf J-2-Triebwerken (Bild: NASA).

Die Treibstofftanks enthalten 371 515 kg LOX und 71 720 kg flüssigen Wasserstoff (LH2). Berechnen Sie für einen Durchmesser der Tanks von 9 m und den Dichten

$$\varrho(\text{LOX}) = 1141 \, \frac{\text{kg}}{\text{m}^3}$$

$$\varrho(\text{LH2}) = 70 \, \frac{\text{kg}}{\text{m}^3}$$

die Höhe der beiden Tanks. Nehmen Sie eine zylindrische Geometrie der Tanks an.

**Endgeschwindigkeit der Saturn V-Rakete im Erdorbit**

Bevor sich die Crew von Apollo 11 auf dem Weg zum Mond machte, brachte die dritte Stufe (S-IVB) der Saturn V sie auf einen Parkorbit um die Erde (Abb. 6). Erst nach 1 1/2 Erdumrundungen feuerte die dritte Stufe erneut und beschleunigte auf eine Geschwindigkeit, die ausreichte, um den Mond zu erreichen.

Bei der Ankunft auf dem Parkorbit in einer Höhe von etwa 169 km hatte Apollo 11 eine Geschwindigkeit von 7390 m/s erzielt (Orloff 2000, S. 103). Dieser Wert bezieht sich jedoch, wie alle bisherigen Geschwindigkeiten, auf die Erdoberfläche. Wenn die Astronauten sich auf den Weg zum Mond machen, ist jedoch die Geschwindigkeit relativ zum Erdmittelpunkt maßgeblich. Der Unterschied zwischen den beiden Betrachtungen liegt in der Rotation der Erde. Die Geschwindigkeit der Erdoberfläche am Startpunkt addiert sich zu der Bahngeschwindigkeit, die die Rakete im Erdorbit erreicht hat. Daher starten die Raketen meist in eine östliche Richtung, also mit der Erddrehung, um die Geschwindigkeit der Erdrotation zu nutzen (Abb. 6).

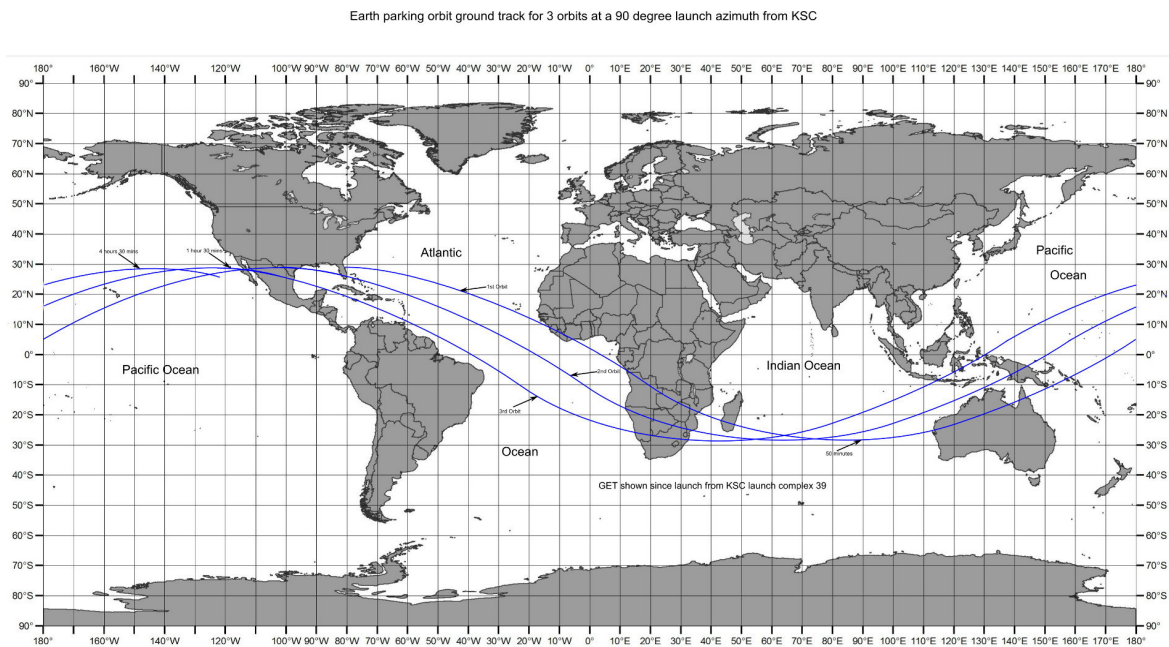

**Abbildung 6:** Diese Grafik zeigt den Parkorbit des Apollo-Raumschiffs vor dem Übergang in den Orbit zum Mond (Bild: NASA/Robin Wheeler, https://history.nasa.gov/afj/launchwindow/lw1.html).

Berechnen Sie die Tangentialgeschwindigkeit der Erdoberfläche am Weltraumbahnhof Cape Canaveral, von wo die Saturn V-Raketen abhoben. Berücksichtigen Sie hierfür die folgenden Parameter.

**Tabelle 2:** Geografische Parameter zur Berechnung der Rotationsgeschwindigkeit der Erde am Weltraumbahnhof Cape Canaveral.

| Parameter | Wert |
|---|---|
| Geografische Breite | +28°35′07″ |
| Äquatorialer Erdradius | 6378 km |

Dieser Wert addiert sich zur oben genannten Bahngeschwindigkeit hinzu. Wie hoch ist die Gesamtgeschwindigkeit?

**Der Weg zum Mond**

Eine wichtige Vorgabe bei der Definition des Orbits zum Mond war, dass es sich um eine freie Rückkehrbahn (engl. free return trajectory) handeln sollte (Zimmer 2009). Sie ist so gewählt, dass das Apollo-Raumschiff nach dem Verlassen der Parkbahn in einem Notfall ohne Antrieb wieder zur Erde zurückkehrt (Abb. 7). Dieses Sicherheitskonzept hat sich während der Apollo 13-Mission bewährt.

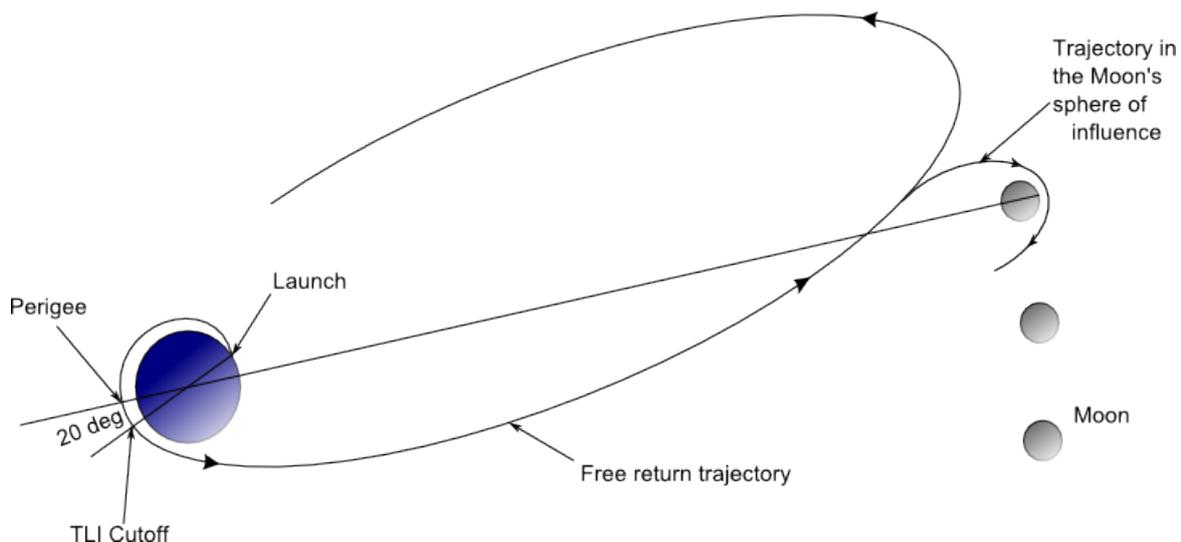

**Abbildung 7:** Diese Grafik zeigt den elliptischen Orbit des Apollo-Raumschiffs zum Mond. Es handelt sich dabei um eine freie Rückkehrbahn (free return trajectory) (Bild: NASA/Robin Wheeler, https://history.nasa.gov/afj/launchwindow/lw1.html).

Um auf diese Bahn zu gelangen (TLI, Trans Lunar Injection), wurde das J-2-Triebwerk ($I_{sp} = w = 4130\,\text{m/s}$) der dritten Raketenstufe noch einmal für $\Delta t = 386,83\,\text{s}$ gezündet. Dadurch wurde das Raumfahrzeug auf eine stark elliptische Bahn geschoben. Durch den gravitativen Einfluss des Monds verließ das Apollo-Raumschiff jedoch schließlich diese Ellipsenbahn und schwenkte in Richtung Mond ein.

Zeigen Sie mit Gl. (16), dass die Beschleunigung auf die Endgeschwindigkeit von 10,8 km/s eher sanft war. Die Masse des Raumfahrzeugs zu Beginn dieser Zündung betrug 139393,02 kg. Bei diesem Manöver wurden insgesamt 71068,40 kg Treibstoff umgesetzt (Orloff 2000, S. 289). Geben Sie die Beschleunigung $a_R(\Delta t)$ als ein Vielfaches der Erdbeschleunigung $g = 9,81\,\frac{\text{m}}{\text{s}^2}$ an.

**Lernkontrolle und Abschluss**

Neben den korrekten Antworten und Ergebnissen, die von einzelnen Schülerinnen und Schülern erzielt wurden, kann eine weitere Lernkontrolle spielerisch erreicht werden. Hier bietet sich beispielsweise *Kahoot!* an. Dies ist eine interaktive Online-Plattform für verschiedene didaktische Methoden. Typischerweise definiert die Lehrperson zunächst ein Quiz, das dann von den Schülerinnen und Schülern mit ihren Smartphones gespielt wird. Als Spielleiter können Sie die Anzahl der korrekten Antworten pro Frage einsehen und grafisch für alle darstellen. Zudem besteht die Möglichkeit, dass Sie nachsehen können, wer richtige und falsche Antworten gegeben hat und darauf gezielt eingehen. So können Sie z. B. nachfragen, warum jemand eine bestimmte Antwort ausgewählt hat.

Ein solches Quiz ist bereits vorbereitet und kann unter folgendem Link abgerufen werden.

<https://create.kahoot.it/share/apollo-11-weg-zum-mond/71135cbc-3779-467f-b813-2ae7481bfe27>

# Lösungen

### „Kugelstoßen" in Schwerelosigkeit

$$S = \mu \cdot w = \frac{\Delta m}{\Delta t} \cdot w = \frac{10\,\text{kg}}{0{,}25\,\text{s}} \cdot 10\,\text{m/s} = 400\,\frac{\text{kg}\cdot\text{m}}{\text{s}^2} = 400\,\text{N}$$

$$\Delta v = w \cdot \ln\left(\frac{m_0}{m_B}\right) = 10\,\text{m/s} \cdot \ln\left(\frac{200\,\text{kg}}{190\,\text{kg}}\right) = 0{,}5\,\text{m/s}$$

### Flugprofil der Saturn V-Rakete

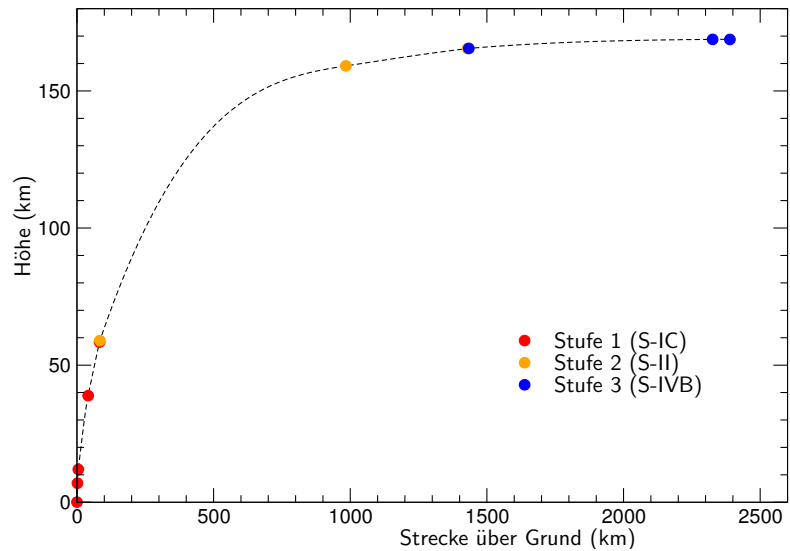

**Abbildung 8:** Flugprofil der Saturn V-Rakete.

Bereits in recht geringer Höhe von etwa 10 km verlässt die Rakete deutlich die senkrechte Ausrichtung. Während der Brennphase der zweiten Stufe wird eine Höhe von 150 km und damit der Parkorbit fast erreicht.

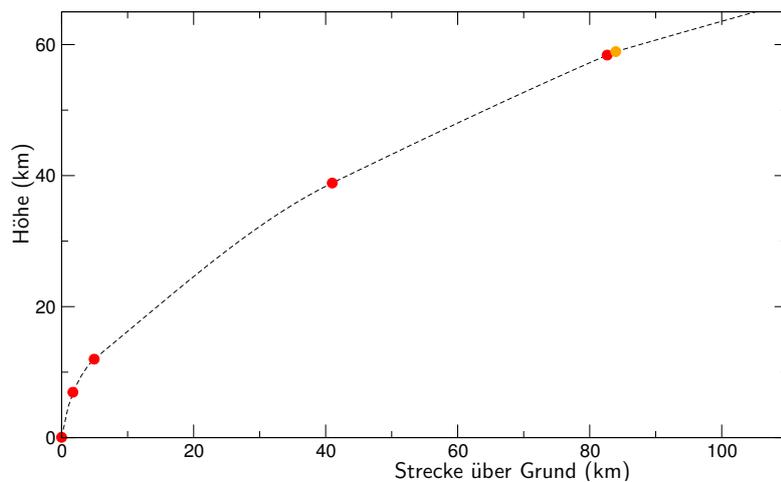

**Abbildung 9:** Flugprofil der Saturn V-Rakete mit gleicher Skalierung beider Achsen.

**Eigenschaften der ersten Stufe der Saturn V-Rakete**

$$\mu = \frac{\Delta m}{\Delta t} = \frac{\Delta m_{\text{LOX}} + \Delta m_{\text{RP}-1}}{\Delta t} = \frac{1481439\,\text{kg} + 632635\,\text{kg}}{168,03\,\text{s}} = 12582\,\text{kg/s}$$

Die erste Stufe setzt pro Sekunde fast 12,6 t Treibstoff um.

$$\Delta v = w \cdot \ln\left(\frac{m_0}{m_B}\right) - g \cdot \Delta t = 3180\,\text{m/s} \cdot \ln\left(\frac{2938315\,\text{kg}}{824241\,\text{kg}}\right) - 9.81\,\frac{\text{m}}{\text{s}^2} \cdot 168,03\,\text{s} = 2394\,\text{m/s}$$

Am Ende der Brenndauer der ersten Stufe hatte die Saturn V bereits eine Geschwindigkeit von knapp 2,4 km/s erreicht.

$$S = \mu \cdot w = 12582\,\text{kg/s} \cdot 3180\,\text{m/s} = 40010760\,\frac{\text{kg} \cdot \text{m}}{\text{s}^2} = 40011\,\text{kN}$$

Wenn man möchte, kann man diese Kraft so darstellen, als ob eine Masse mit dieser Gewichtskraft zum Erdboden gezogen wird. Der Schub ist damit äquivalent zu einer Gewichtskraft entsprechend einer Masse von mehr als 4000 t. Im Ingenieursbereich wird diese anschauliche Darstellung gerne benutzt und von einem Schub von mehr als 4000 t gesprochen.

**Treibstoff der zweiten Stufe der Saturn V-Rakete**

$$V = \pi \cdot r^2 \cdot h$$
$$\Leftrightarrow h = \frac{V}{\pi \cdot r^2} = \frac{m}{\pi \cdot r^2 \cdot \varrho}$$

$$h_{\text{LH2}} = \frac{71720\,\text{kg}}{\pi \cdot (4,5\,\text{m})^2 \cdot 70\,\text{kg/m}^3} = 16,1\,\text{m}$$

$$h_{\text{LOX}} = \frac{371515\,\text{kg}}{\pi \cdot (4,5\,\text{m})^2 \cdot 1141\,\text{kg/m}^3} = 5,1\,\text{m}$$

In dieser einfachen Betrachtung haben die beiden Tanks eine Höhe von 21,2 m.

## Endgeschwindigkeit der Saturn V-Rakete im Erdorbit

Auf der rotierenden Erde verlaufen alle Breitenkreise parallel zueinander. Insbesondere ist der Breitenkreis zu einer geografischen Breite $\phi$ parallel zum Äquator (Abb. 10). Der Umfang eines Breitenkreises $U_\phi$ ist stets geringer als der Erdumfang am Äquator $U$.

Mit Abb. 11 erkennt man, dass der Radius eines Breitenkreises $R_\phi$ über den Breitengrad $\phi$ mit dem Erdradius $R$ verknüpft ist. Es gilt:

$$R_\phi = R \cdot \cos(\phi)$$

Und somit:

$$U_\phi = 2 \cdot \pi \cdot R_\phi = 2 \cdot \pi \cdot R \cdot \cos(\phi) = U \cdot \cos(\phi)$$

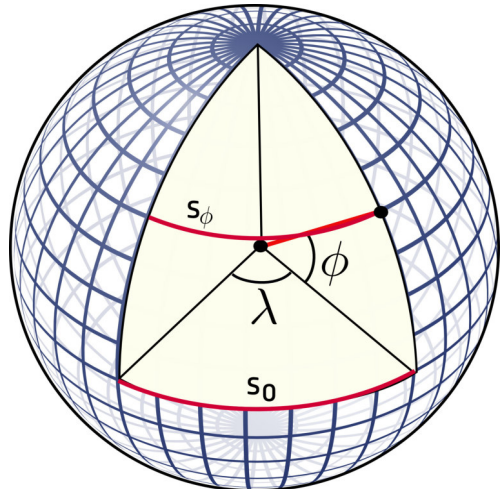

**Abbildung 10:** Geografische Länge und Breite der Erde. Das Segment $s_\phi$ bei der geografischen Breite $\phi$ verläuft parallel zum Segment $s_0$ am Äquator.

Die Erde rotiert innerhalb einer Zeitspanne $\tau = 86164\,\text{s}$ um die eigene Achse, die man den siderischen Tag nennt. Diese ist etwas kleiner als der bürgerliche Tag, der sich nach dem Sonnenstand richtet.

Die Geschwindigkeit $v_\phi$, mit der sich die Erdoberfläche an der geografischen Breite $\phi$ bewegt, ist demnach der Umfang geteilt durch die Zeit für eine Umrundung.

$$v_\phi = \frac{U_\phi}{\tau} = \frac{U \cdot \cos(\phi)}{\tau} = \frac{2 \cdot \pi \cdot R \cdot \cos(\phi)}{\tau} = \frac{2 \cdot \pi \cdot 6378\,\text{km} \cdot \cos(28,5823°)}{86164\,\text{s}} = 408\,\text{m/s}$$

Die Geschwindigkeit im Parkorbit beträgt demnach $v_{\text{end}} + v_\phi = 7390\,\text{m/s} + 408\,\text{m/s} = 7798\,\text{m/s}$.

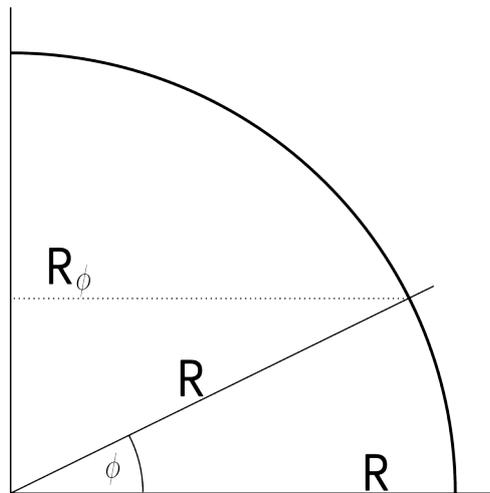

**Abbildung 11:** Der Radius $R_\phi$ des Segments $s_\phi$ bei der geografischen Breite $\phi$ ist über den Kosinus der geografischen Breite mit dem Erdradius $R$ verknüpft.

## Der Weg zum Mond

$$
\begin{aligned}
a_R(t) &= \frac{\mu \cdot w}{m_0 - \mu \cdot \Delta t} \\[6pt]
&= \frac{\frac{\Delta m}{\Delta t} \cdot w}{m_0 - \frac{\Delta m}{\Delta t} \cdot \Delta t} \\[6pt]
&= \frac{\frac{71068{,}4\,\text{kg}}{346{,}83\,\text{s}} \cdot 4130\,\text{m/s}}{139393{,}02\,\text{kg} - 71068{,}4\,\text{kg}} \\[6pt]
&= \frac{846271{,}9\,\frac{\text{kg}}{\text{m}\cdot\text{s}^2}}{68324{,}62\,\text{kg}} = 12{,}39\,\frac{\text{m}}{\text{s}^2} = 1{,}26 \cdot g
\end{aligned}
$$

**Endnoten**

[a]Спутник, russisch: Begleiter, Satellit

[b]Der Name „Apollo" wurde vom damaligen Leiter des NASA-Raumfahrtprogramms, Abe Silverstein nach dem Gott Apoll gewählt, der nach der griechisch-römischen Mythologie den Sonnenwagen zieht (Sapienza 2009).

# Literatur

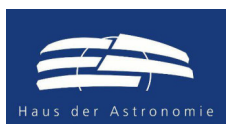
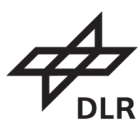
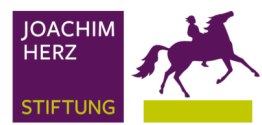